\documentstyle[twoside,fleqn,espcrc2,epsfig]{article}

\newcommand{\lsim}{\mathrel{\rlap{\lower4pt\hbox{\hskip1pt$\sim$}}
    \raise1pt\hbox{$<$}}}         
\newcommand{\gsim}{\mathrel{\rlap{\lower4pt\hbox{\hskip1pt$\sim$}}
    \raise1pt\hbox{$>$}}}         
\newcommand{\esim}{\mathrel{\rlap{\raise2pt\hbox{$\sim$}}
    \lower1pt\hbox{$-$}}}         

\begin{document}

\thispagestyle{empty}

\rightline{MPI-PhT/97-78}
\rightline{hep-ph/9711461}
\rightline{November 1997}

\vspace{3pc}

\centerline{\Large \bf Neutralino relic density including 
coannihilations}

\vspace{3pc}

\centerline{\large Paolo Gondolo}
\smallskip
\centerline{\em Max-Planck-Institut f\"ur Physik (Werner-Heisenberg-Institut)}
\centerline{\em F\"ohringer Ring 6, 80805 M\"unchen, Germany}
\centerline{\tt gondolo@mppmu.mpg.de}

\bigskip

\centerline{\large Joakim Edsj{\"o}}
\smallskip
\centerline{\em Department of Theoretical Physics, Uppsala University}
\centerline{\em P.O. Box 803, SE-751 08 Uppsala, Sweden}
\centerline{\tt edsjo@teorfys.uu.se}

\vspace{3pc}

\centerline{\large \bf Abstract}
\bigskip

  We give an overview of our precise calculation of the relic density of the
  lightest neutralino, in which we included relativistic Boltzmann averaging,
  subthreshold and resonant annihilations, and coannihilation processes with
  charginos and neutralinos.

\vfill

\begin{center}
{ \em
Talk presented by Paolo Gondolo at ``Topics in Astroparticle and Underground
Physics (TAUP) 97,'' Laboratori Nazionali del Gran Sasso, Italy, 7--11
September 1997.
}
\end{center}

\vspace{2pc}
\eject
\setcounter{page}{1}

\title{Neutralino relic density including coannihilations\thanks{Presented by
    P. Gondolo.}}

\author{P. Gondolo\address{Max Planck Institute for Physics,
        F\"ohringer Ring 6, 80805 M\"unchen, Germany}
        and 
        J. Edsj\"o\address{Department of Theoretical Physics, 
          Uppsala University,
          P.O. Box 803, SE-751 08 Uppsala, Sweden}%
        \thanks{Present address: Center for Particle Astrophysics, University
          of California, Berkeley, 301 Le Conte Hall, Berkeley, 
          CA 94720-7304, U.S.A.}
        }
       
\begin{abstract}
  We give an overview of our precise calculation of the relic density of the
  lightest neutralino, in which we included relativistic Boltzmann averaging,
  subthreshold and resonant annihilations, and coannihilation processes with
  charginos and neutralinos.
\end{abstract}

\maketitle

\section{INTRODUCTION}

The lightest neutralino is one of the most promising candidates for the dark
matter in the Universe. A linear combination of the superpartners of the
neutral gauge and Higgs bosons, it is believed to be the lightest stable
supersymmetric particle in the Minimal Supersymmetric extension of the Standard
Model (MSSM). 

In the near future, high precision measurements of the dark matter density may
become possible from high resolution maps of the cosmic microwave background,
and this may constrain supersymmetry. It is therefore of great interest to
calculate the relic density of the lightest neutralino as accurately as
possible.

As a major step towards a complete and precise calculation valid for all
neutralino masses and compositions, we included for the first time \cite{EG}
all concomitant annihilations (coannihilations) 
between neutralinos and charginos, properly
treating thermal averaging in presence of thresholds and resonances in the
annihilation cross sections.


\section{FORMALISM}
\label{sec:Boltzmann}

Consider coannihilation of $N$ supersymmetric particles with masses $m_i$ and
statistical weights $g_i$ (first studied in ref.~\cite{Salati}).  Normally,
all heavy particles have time to decay into the lightest one, which we assume
stable. Its final abundance is then simply described by the sum of the
densities $ n= \sum n_{i}.$ When the scattering rate of supersymmetric
particles off the thermal background is much faster than their annihilation
rate, $n$ obeys the evolution equation \cite{GriestSeckel}
\begin{equation} \label{eq:Boltzmann2}
  \frac{dn}{dt} =
  -3Hn - \langle \sigma_{\rm{eff}} v \rangle 
  \left( n^2 - n_{\rm{eq}}^2 \right)
\end{equation}
with effective annihilation cross section
\begin{equation} \label{eq:sigmaveffdef}
  \langle \sigma_{\rm{eff}} v \rangle = 
  \frac{A}{n_{\rm{eq}}^2} \, .
\end{equation}
The numerator $A$ is the total annihilation rate per unit volume
at temperature $T$, and $n_{\rm eq}$ is the total equilibrium density.
Under the assumption of Boltzmann statistics (a good approximation for $T \lsim
m_i$), we obtain~\cite{EG}
\begin{equation} \label{eq:neq2}
  n_{\rm eq} = 
  \frac{T}{2\pi^2} \sum_i g_i m_{i}^2
  K_{2}\!\left( \frac{m_{i}}{T}\right)
\end{equation}
and
\begin{equation} \label{eq:A}
  A = \frac{T}{16 \pi^4} \int_{4m_\chi^2}^\infty \!ds
  \sqrt{s-4m_\chi^2}
  K_{1}\!\left( \frac{\sqrt{s}}{T}\right) \, 
  W(s) .
\end{equation}
Here $K_i(x)$ is the modified Bessel function of the second kind of 
order $i$, and
\begin{equation} \label{eq:Wijcross}
  W(s) = \sum_{ij}  
  \frac{\lambda(s,m_i^2,m_j^2)}{2\sqrt{\lambda(s,m_\chi^2,m_\chi^2)}}
  \, g_i g_j \sigma_{ij} 
\end{equation}
where $\lambda(x,y,z) = x^2+y^2+z^2-2xy-2xz-2yz$.

$W(s)$ is a Lorentz invariant annihilation rate per unit volume in which
coannihilations appear as thresholds at $\sqrt{s}$ equal to the sum of the
masses of the coannihilating particles.  The independence of $W(s)$ on
temperature is a remarkable calculational advantage in presence of
coannihilations: in fact it can be tabulated in advance, before taking the
thermal average and solving the density evolution equation.

Eqs.~(\ref{eq:sigmaveffdef}-\ref{eq:Wijcross}) generalize the result of
Gondolo and Gelmini~\cite{GondoloGelmini} to coannihilations.


\section{RESULTS}
\label{sec:Results}

To explore a significant fraction of the MSSM parameter space \cite{haberkane},
we keep the number of theoretical relations among the parameters to a minimum.
We assume GUT relations for gaugino masses, keep only the top and bottom
trilinear soft supersymmetry-breaking parameters, and use a single mass
parameter for the diagonal entries in the sfermion mass matrices at the weak
scale. We perform many different scans in parameter space, some general and
some specialized to interesting regions.  We keep only models that satisfy the
experimental constraints on the $Z^0$ width, on the $b \rightarrow s \gamma$
branching ratio, and on superpartner and Higgs boson masses.

\begin{figure}[hb]
  \vbox to 2.2in{\vskip-0.3in\epsfig{file=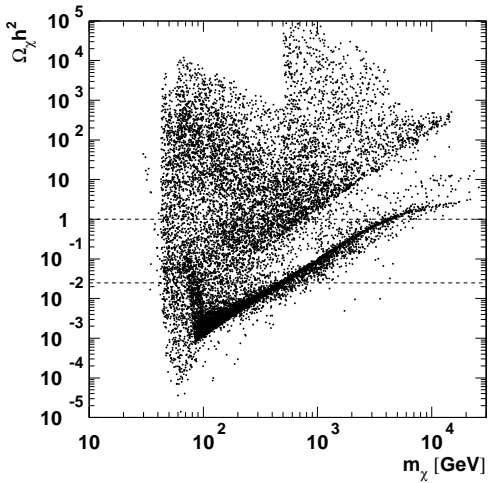,width=0.45\textwidth}}
  \caption{
    Neutralino relic density including neutralino and chargino
    coannihilations versus neutralino mass.  The horizontal lines
    bound the cosmologically interesting region $0.025 < \Omega_\chi
    h^2 <1$.}
  \label{fig:oh2vsmx}
\end{figure}

We obtain analytic expressions for the many Feynman diagrams contributing to
the two-body cross sections at tree level for
neutralino-neutral\-ino, neutralino-chargino and chargino-chargino
annihilation. Then for each set of model parameters, we sum over particle
polarizations and over initial and final states numerically and tabulate the
annihilation rate $W(s)$. Thermal averaging and integration of the density
equation finally give the neutralino relic density $\Omega_\chi h^2 = m_\chi
n_0/\rho_{\rm crit}$ in units of the critical density $\rho_{\rm crit}$.

Fig.~\ref{fig:oh2vsmx} shows the neutralino relic density $\Omega_\chi h^2$
versus the neutralino mass $m_\chi$. Each point represents a set of model
parameters. It should be kept in mind that bands and holes in the point
distributions are mere artifacts of our sampling in parameter space.

The horizontal lines limit the cosmologically interesting region where the
neutralino can constitute most of the dark matter in galaxies without violating
the constraint on the age of the Universe. We take it to be $0.025 <
\Omega_{\chi} h^2 < 1$.

\begin{figure}[t]
  \vbox to 2.2in{\vskip-0.3in\epsfig{file=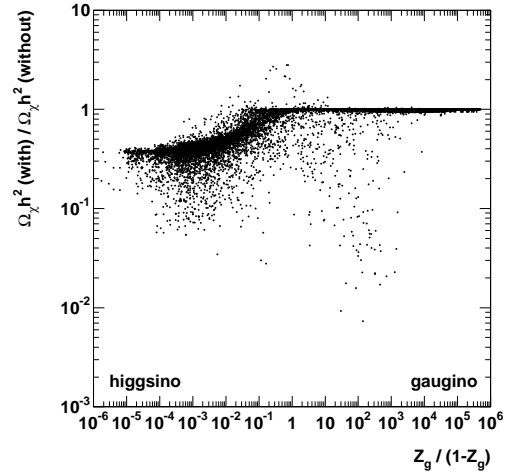,width=0.45\textwidth}}
  \caption{
    Ratio of the neutralino relic density with and without
    neutralino and chargino coannihilations versus 
    neutralino composition. }
  \label{fig:ratiovsmx}
  \vskip-\baselineskip
\end{figure}

The effect of coannihilations on the neutralino relic density
is summarized in Fig.~\ref{fig:ratiovsmx}, 
which shows the ratio of
the neutralino relic density with and without coannihilations versus the
neutralino composition $Z_g/(1-Z_g)$ (here $Z_g$ is the gaugino fraction). 

Coannihilation processes are important not only for light
higgsino-like neutralinos, as pointed out
before in approximate calculations~\cite{MizutaYamaguchi}, but also for heavy
higgsinos and for mixed and gaugino-like neutralinos.  Indeed, coannihilations
should be included whenever $|\mu| \lsim 2 |M_1|$, independently of the
neutralino composition. When $|\mu| \sim |M_1|$, coannihilations can increase
or decrease the relic density in and out of the cosmologically interesting
region.

\begin{figure}[t]
  \vbox to 2.2in{\vskip-0.3in\epsfig{file=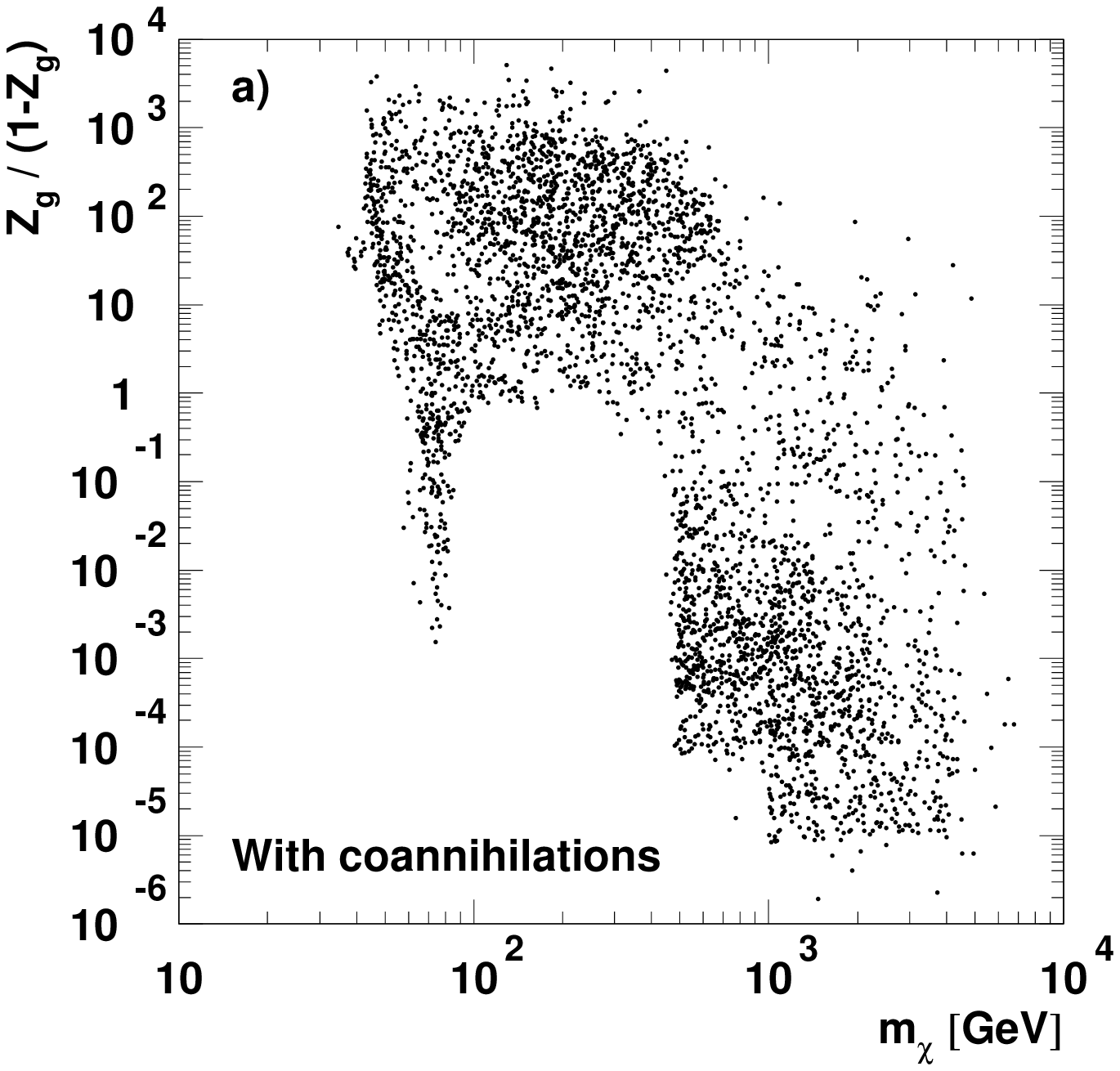,width=0.45\textwidth}}
  \vbox to 2.4in{\epsfig{file=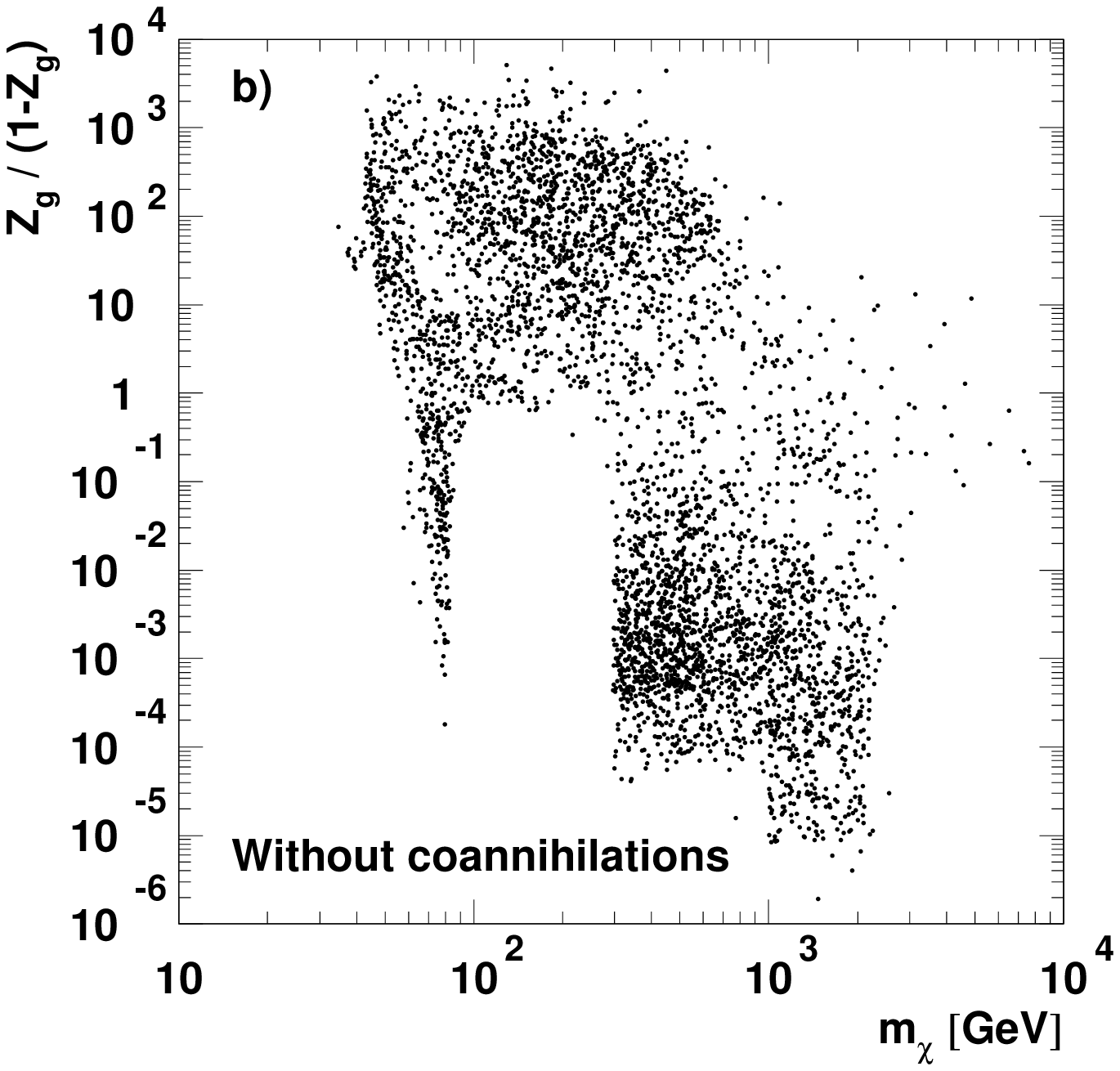,width=0.45\textwidth}}
  \caption{Neutralino masses $m_\chi$ and compositions $Z_g/(1-Z_g)$
    for cosmologically interesting models (a) with and (b) without
    coannihilations.}
  \label{fig:cosmregion}
\end{figure}

Fig.~\ref{fig:cosmregion} shows the cosmologically interesting region in the
neutralino mass--composition plane, before and after including coannihilations.
This region is limited to the left by accelerator
constraints, to the right by $\Omega_\chi h^2 < 1$, and below and above by
incompleteness in our survey of parameter space, except for the hole at 85--450
GeV where $\Omega_\chi h^2 < 0.025$.

The main effect of coannihilations is to shift the higgsino region to higher
masses. In particular the cosmological upper bound on the neutralino mass
changes from 3 to 7 TeV.  Differently from previous approximate
results~\cite{MizutaYamaguchi}, there is a cosmologically interesting
window of light higgsino-like neutralinos with masses around 75 GeV.

We conclude that if coannihilations are properly included, the neutralino is a
good dark matter candidate whether it is light or heavy, and whether it is
higgsino-like, mixed or gaugino-like.

\vskip\baselineskip

P.~Gondolo thanks P.~Salati for bringing ref.~\cite{Salati} to his attention.


\end{document}